\documentclass[preprint,epsf]{aastex}
\input epsf

\received{}
\accepted{}
\accepted{}
\ccc{}
\cpright{}{}

\slugcomment{Submitted to AJ (March 8, 2000)}
\shorttitle{[Fe/H] of Red Giants in the Bulge}
\shortauthors{Ram\'{\i}rez et al.}

\begin{document}

\title{Metallicity of Red Giants in the Galactic Bulge from Near-Infrared
Spectroscopy} 

\vspace{10mm}
\author{Solange V. Ram\'{\i}rez\altaffilmark{1,2}, Andrew Stephens, 
Jay A. Frogel\altaffilmark{1}, D. L. DePoy\altaffilmark{1}}
\affil{Department of Astronomy, The Ohio State University}

\vspace{10mm}
\altaffiltext{1}{Visiting Astronomer, Cerro Tololo Inter-American
Observatory.
CTIO is operated by AURA, Inc.\ under contract to the National Science
Foundation.}
\altaffiltext{2}{US Gemini Fellow}

\begin{abstract}
We present $K-$band spectra of more than 110 M giants in Galactic bulge fields 
interior to $-$4 degrees and as close as 0.2 degrees of the Galactic Center. 
From the equivalent widths of three features in these spectra, EW(Na),
EW(Ca), and EW(CO) we calculate [Fe/H] for the stars with a calibration derived
from globular clusters \citep{ste00}.
The mean [Fe/H] for each field is in good agreement with the results from
\citet{fro99} based on the slope of the giant branch method.
We find no evidence for a metallicity gradient along the minor 
or major axes of the inner bulge ($R < $ 0.6 kpc). 
A metallicity gradient along the minor axis, found earlier, arises when 
fields located at larger galactic radius are included. 
However, these more distant fields are located outside of the infrared bulge 
defined by the COBE/DIRBE observations.

We compute the [Fe/H] distribution for the inner bulge and find a mean value of
-0.21 dex with a full width dispersion of 0.30 dex, close to the values 
found for Baade's Window (BW) by \citet{sad96} and to a theoretical
prediction for a bulge formed by dissipative collapse \citep{mol00}.
\end{abstract}

\section{Introduction}

Baade's Window (BW, $l=1^{\circ}, b=-4^{\circ}$) is the most studied 
region in the Galactic bulge.
In the late 1970s, \citet{whi78} demonstrated that the integrated 
optical spectrum from BW closely resembles that from bulges of spiral 
galaxies and from moderate luminosity E and S0 galaxies.
At the same time \citet{fro78} found that the near infrared light from these 
galaxies is dominated by cool giant stars. 
Shortly thereafter, Blanco and his collaborators determined that BW 
contains an unusually high percentage of middle and late type M giants 
compared to other regions of the Galaxy \citep{bla84}. 
Detailed studies of the M giants in BW revealed that they have photometric 
and spectroscopic properties significantly different from those of M 
giants in the field \citep{fro87,fro88,fro90,ric83,ric88,ter90,mcw94}.

The accurate determination of stellar metallicities is essential for 
constraining models of 
star formation and chemical evolution in the bulge. 
\citet{fro90} used $JHK$ colors and CO and {$\rm H_{2}O$} photometric 
indices to determine metallicities of stars at latitudes between 
$-3^{\circ}$ and $-12^{\circ}$ along the minor axis. 
\citet{tie95} used the relation between the slope of the 
upper giant branch and [Fe/H] \citep{kuc95a,kuc95b} to estimate 
metallicities for the same stars studied by \citet{fro90}.
\citet{tys91} used Washington photometry on stars in similar fields 
for the same purpose. 
These three studies agreed that there is a small metallicity gradient 
along the minor axis of the bulge, with values ranging from -0.04 dex/deg 
\citep{fro90} to -0.09 dex/deg \citep{tys91}.
Finally, \citet{min95} discussed evidence for a 
metallicity gradient in the Galactic bulge based on compiled observations of 
ten fields, eight of them exterior to BW. 

Stars in most of the inner bulge ($ |b| \leq 3^{\circ}$) 
can be studied only with near infrared observations because of high 
reddening and extinction.  Recently, \citet{fro99} have 
studied 11 fields in the inner Galactic bulge using $JHK$ photometry.
Seven of these fields are on the minor axis; five are at a latitude of
$-1.3^{\circ}$ parallel to the major axis. They estimated the reddening
of each field from their CMDs and the mean metallicity of each
field with the giant branch slope method.
They combined their results with those of \citet{tie95} and  
derived a gradient of -0.064 $\pm$ 0.012 dex/degree in the range 
$ -0.2^{\circ} \leq b \leq -10.25^{\circ}$ along the minor axis.

Our main goal is to obtain independent values for the metallicity of the 
stars in the same inner bulge fields studied by \citet{fro99} but with 
spectroscopic techniques. 
We use the strength of three absorption features
present in the $K-$band of cool stars: Na, Ca, and CO.
The calibration is based on similar observations of giants in globular 
clusters by \citet{ste00}.  

\section{Observations and Data Reduction}

\subsection{Observations}

The observations for this paper were obtained on the 4m Blanco telescope at 
Cerro Tololo Inter-American Observatory (CTIO) during three observing runs
with two instruments: the Ohio State InfraRed Imager and Spectrograph 
\citep[OSIRIS; $R = 1380 $;][]{dep93}, and the CTIO
InfraRed Spectrograph \citep[IRS; $R = 1650, 4830 $;][]{dep90}. 
OSIRIS used a 256 $\times$ 256 NICMOS3 detector and IRS has
a 256 $\times$ 256 InSb array.  Spectral coverage was between 2.19 $\mu$m 
and 2.32 $\mu$m with both instruments. 
Table 1 gives a brief log of the observing runs.

Program stars were selected from the color magnitude diagrams of 11 fields 
interior to -4 degrees and as close as 0.2 degrees to the Galactic Center 
\citep{fro99}. All of the observed stars are at or near the top of 
the red giant branch of each field. 
Based on their location in the CMDs we believe their probability of 
membership in the bulge is high.
Fields are designated as in \citet{fro99}.
Our sample also includes 14 stars in BW from \citet{fro87} and \citet{ter91}.
Table 2 lists our final sample of stars. Column 1 gives the star's name,
column 2 the date of observation (year, month, day), column 3 the spectral 
resolution, column 4 the estimated S/N ratio per pixel in the spectrum, 
columns 5 and 6 the observed $K$ magnitude and $J-K$ colors 
\citep[][and unpublished]{fro99}, columns 7 and 8 the absolute $K$magnitude 
and dereddened $J-K$ colors (see section 3.2), 
and columns 9, 10 and 11 the equivalent widths of the Na, Ca, and CO features.
The tabulated uncertainties in these three quantities are simply the ratio of 
the mean continuum to the rms noise present in the defined continuum regions 
(see section 3.1). 

\subsection{Data Reduction}

The data acquisition and reduction were similar for all the spectra.
Both instruments, OSIRIS and IRS, were used first in imaging mode 
to acquire the star in the slit. 
We took several spectra ($\sim$ 10) with the star stepped along the slit. 
This was done to estimate the sky levels in the exposures, to compensate
for bad pixels, and to aid in the removal of the fringes present in the 
OSIRIS data which are typically $\sim$6 \% (peak-to-peak) of the continuum.
A star of spectral type A or B was observed as close
to the object star's airmass as possible to correct for telluric
absorption features. Such stars have no significant spectral
features in the wavelength region we observed.
After we averaged the multiple spectra of a program star 
and divided by the average spectrum of the nearby atmospheric standard star, 
the fringes cancelled to $<$1\%.
OSIRIS and IRS at $\lambda / \Delta\lambda$ = 1650 can cover the 
entire relevant spectral region (2.19$\mu$m -- 2.34$\mu$m) with a single 
grating setting; with IRS at $\lambda / \Delta\lambda$ = 4830, however,
we observed at three overlapping grating settings (2.191$\mu$m -- 2.242$\mu$m,
2.238$\mu$m -- 2.290$\mu$m, and 2.285$\mu$m -- 2.337$\mu$m) to cover the
desired wavelength range. A number of stars were observed at both grating
settings as a quality control check. We will discuss results of this check
later. 

We used Image Reduction and Analysis Facility (IRAF\altaffilmark{5})
software for data reduction.
\altaffiltext{5}{The IRAF software is distributed by the National
Optical Astronomy Observatories under contract with the National Science
Foundation}
The reduction process consisted of flat fielding the individual spectra
with dome flats and
sky subtraction using a sky frame made by a median combination of all data
frames of the object.
We replaced bad pixels (dead pixels and cosmic ray hits) by an
interpolated value computed from neighboring pixels in the dispersion
direction.
The OSIRIS frames were then geometrically transformed to correct the
curvature of the slit induced by the grating (maximum correction $\sim $
2 pixels). The induced curvature in IRS frames was insignificant ($<$
1 pixel) over the region of the array that was used.
Geometric transformations for OSIRIS were derived from night sky 
emission lines.
We extracted the individual spectra along an aperture of 7 pixels
using the APSUM package in IRAF and did a further sky
subtraction by using regions on either side of the aperture.
The final spectrum is an average of the extracted spectra.
We divided the spectra of object stars by an 
early type atmospheric standard star, observed and reduced in 
the same way, 
to remove telluric absorption features and multiplied by a 10,000 K 
blackbody to put the spectra on a ${\rm F_{\lambda}}$ scale. 
The temperature of the blackbody approximately corresponds to the
average temperature of our atmospheric standard stars.
The maximum effect from the difference between the standard star temperature
and the temperature of the adopted 10,000 K blackbody is a $\sim$1.5\%
tilt in the continuum slope, which is
insignificant to the equivalent width measurements.

We used OH air-glow lines \citep{oli92} to obtain wavelength solutions.
For OSIRIS and IRS at R = 1650 spectra, we also included the 
${\rm ^{12}CO}$(2,0) bandheads near zero radial 
velocity to compute the wavelength calibration since there are
no OH air-glow lines present at the red end of the $K$-band spectra.
OH air-glow lines were present only in the first grating setting
(2.191$\mu$m -- 2.242$\mu$m);
a second-order wavelength solution was derived there.
Since the grating is the same for the three grating settings and the
grating angle differences are small,
the first and second order terms of the first grating setting solution
should be the same for all three observed grating settings.
After
applying those terms to the second (2.238$\mu$m -- 2.290$\mu$m) and
third (2.285$\mu$m -- 2.337$\mu$m)
grating settings, only the zero order term is unknown. This
zero order term is just a shift that was computed from the lines
present in the overlapping regions.
We connected the three grating settings, after wavelength calibration,
by averaging the overlapping regions.

We shifted both the OSIRIS and IRS spectra in
wavelength to correct for radial velocity differences; the $
^{12}$CO(2,0) bandhead was fixed at 2.293 $\mu$m. This shift is needed to
ensure the relative accuracy and consistency of the equivalent widths of 
atomic and molecular features. A sample of the final normalized spectra 
are shown in Figures 1 on an ${\rm F_{\lambda}}$ scale.
Only the brightest stars for each field are shown in the Figures.
The whole database is available on the anonymous FTP site of the OSU Astronomy
Department (ftp to ftp.astronomy.ohio-state.edu, login as anonymous, change to
directory pub/solange, and get the file bulge\_spec.tar.gz). 

\section{Analysis}
 
\subsection{Equivalent Widths of Atomic and Molecular Absorption Features}

Stellar photospheric absorption features in our spectra were identified
from the wavelengths of the lines in \citet{kle86}.
The strongest features in our data are lines of
Na I and Ca I, and the (2,0) bands of $ ^{12}$CO.
The equivalent widths of these features were measured with respect to a
continuum level defined as the best first-order fit for bands free of 
spectral lines near the features.
The band passes adopted for the features and continuum are given in 
Table 3 and the features themselves are shown in Figure 1.
These bandpasses are identical to those used to measure the giants
stars in globular clusters \citep{ste00}.
The measured equivalent widths for Na I, Ca I, and $ ^{12}$CO(2,0) for 
our program stars are listed in Table 2.

To estimate the {\it formal} uncertainties, we assume that the noise is
dominated by photon statistics and that
$ {\rm \sigma_{line} \sim \sigma_{cont}} $.
The uncertainty in the measurement of each feature (in \AA) is given by:
$$ {\rm \sqrt{2N_{pixels}} \times dispersion \times \sigma_{cont}} $$
\noindent
where $ {\rm N_{pixels}} $ is the number of pixels contained
within the defined feature band, dispersion is measured in \AA \  per
pixel, and ${\rm \sigma_{cont}} $ is the rms noise per pixel of the
fitted continuum.
Uncertainties listed in Table 2 were calculated using this formula.
These values are really lower limits as they provide no estimate for any
systematic errors that may exist in the data.
More realistic estimates of the uncertainties are computed below, using
differences found in measurements taken with different instruments and 
spectral resolutions.

Each feature at each resolution has its own calculated rms noise.
The estimated signal-to-noise ratio
that appears in Table 2 is the ratio of the continuum level to the
standard deviation of each spectrum.
The standard deviation of each spectrum is the
quadratic average of the calculated rms noise of each feature
(${\rm \sigma_{cont} =
\sqrt{[\sigma^{2}_{cont}(Na) + \sigma^{2}_{cont}(Ca) +
\sigma^{2}_{cont}(CO)]/3}}$).

There are 13 stars with IRS spectra at both spectral resolutions 1650 \& 4830. 
Also, there are 8 stars with spectra taken with OSIRIS at R=1380 and IRS at 
R=1650. 
The average differences and standard deviation of Na, Ca, and CO equivalent 
widths measured at both resolutions are listed in Table 4.
The last column of Table 4 is the average formal error from Table 2.
The mean differences of the equivalent widths measured 
at different resolutions and taken with different instruments are 
negligible at the one sigma level. 
Thus, when we have more than one observation we will use the 
average of the equivalent widths measured at different resolutions.
The standard deviations listed in Table 4 are also an indicator of potential
systematic uncertainties in the data; we consider these values to be a 
better
estimate of the true uncertainties in the respective equivalent widths than 
the formal errors. The total uncertainties are computed as the average
of the standard deviations listed in Table 4, and are 0.38 for EW(Na), 
0.87 for EW(Ca) and 1.7 for EW(CO).

\subsection{Reddening} 

We estimated the extinction and reddening to each star using the same
technique as \citet{fro99}. Specifically, we assumed that the color of the
upper giant branch in each field was the same as that in BW:
\begin{equation}
(J-K)_{0} = -0.113 K_{0} + 2.001
\end{equation}
where $(J-K)_{0}$ is the dereddened $J-K$ color and $K_{0}$ is the 
dereddened $K$ magnitude. Further, we assumed the relation between 
extinction and reddening found by \citet{mat90}:
\begin{equation}
A_{K} = 0.618 E(J-K).
\end{equation}
The reddening is estimated by calculating the shift in $K$ and $J-K$
along the reddening vector to force each star to fall on the BW giant
branch. $M_{K_{0}}$ is computed assuming that all stars are located at 
a distance of 8 kpc. Dereddened photometric indices, 
$M_{K_{0}}$ and $(J-K)_{0}$, are listed in Table 2.

The photometric uncertainties are estimated to be about 0.04 or 0.05
magnitudes \citep{fro99} for $K$ and $(J-K)$. The uncertainties of the 
dereddened photometric indices should also include the uncertainties caused by
the differential reddening present in each field and the assumption that all 
stars are located at the same distance.
\citet{fro99} found that the amount of scatter due to differential reddening
is proportional to the average reddening for each field. Using eq. (3) from
\citet{fro99} we estimate that the scatter due to differential reddening
implies an scatter in $(J-K)_{0}$ of 0.30 mag for the c fields and 0.12 mag for
the g fields, and a scatter in $M_{K_{0}}$ of 0.2 mag for the c fields and
0.1 mag for the g fields and BW. 
The maximum dispersion in magnitude due to spread along the line of sight is
0.2 mag \citep{fro90} for fields at galactic latitude less than 4$^{\circ}$. 
So, the scatter in $M_{K_{0}}$ including both the effects of 
differential reddening and
dispersion along the line of sight is 0.30 mag for the c fields and 0.23 mag
for the g fields and BW.

\section{Results}

\subsection{Dependence on Luminosity}

Figure 2 shows the dependence of the equivalent widths of Na, Ca, and 
CO on $M_{K_{0}}$. 
These plots resemble CMDs, since the EW(Na), EW(Ca), and EW(CO)
may depend on both effective temperature and luminosity in addition to
metallicity.

There is a considerable amount of scatter in Figure 2.
We computed the standard deviation of EW(Na), EW(Ca), and EW(CO) in two
narrow ranges of $M_{K_{0}}$, listed in Table 5, to minimize any spread
that might arise from an $M_{K_{0}}$ dependence of the indices. 
In all cases the standard
deviation is greater than the total uncertainties of the equivalent widths
(see Sec. 3.1).
Therefore, part of the scatter is real and can be understood as a spread
in metallicity in our sample of bulge stars.
Note that we assume all stars are of closely similar age so that only [Fe/H]
differences will cause a spread in color or EW at a given $M_{K_{0}}$.
If the observed scatter is the quadratic addition of the scatter due to 
differences in metallicity and the scatter due to uncertainties in 
the data, then the scatter due to metallicity is 0.9 \AA ~for EW(Na), 
0.4 \AA ~for EW(Ca), and 1.8 \AA ~for EW(CO).

There is a statistically significant slope of -0.6 \AA/mag in EW(CO) vs. 
$M_{K_{0}}$.
However, this probably reflects the dependence of both EW(CO) and $M_{K_{0}}$
on $(J-K)_{0}$ color \citep[see ][]{joh66,ram97}.
In particular, as giants evolve they increase in luminosity and decrease
in effective temperature. As the effective temperature decreases, the CO
opacity, and hence the strength of the CO lines, increases. 
For example, there is a slope of $\sim$ -0.5 \AA/mag in the EW(CO) vs. 
$M_{K_{0}}$ relation for the field giants, assuming that the giant branch 
of BW (eq. 1) is similar to the giant branch of field giants.
This is very similar to the slope of -0.6 \AA/mag we observe.
This suggests that the slope we observe in EW(CO) vs. $M_{K_{0}}$ in 
Figure 2 can be explained by the dependence of both EW(CO) and $M_{K_{0}}$
on effective temperature or $(J-K)_{0}$ color.

There is no obvious relation between EW(Na) and EW(Ca) with respect to
$M_{K_{0}}$. Our previous study of field giants \citep{ram97} indicates 
that such a relation should exist (for the same reasons as for CO). 
But, the scatter is too high in the graphs of EW(Na) and EW(Ca) 
vs. $M_{K_{0}}$ to find a relationship between those indices.

\subsection{Metallicity using Globular Cluster Giants}

\citet{ste00} have established an [Fe/H] scale for Galactic globular
clusters based on medium resolution (1500-3000) infrared $K$ band spectra
of the brightest stars in 15 clusters.  The technique uses the same 
absorption features as we use here: Na, Ca, and CO.
Indeed, many of their spectra were obtained with the identical instrument
set up and on the same nights as the spectra analyzed here.
Their calibration is derived from spectra of more than 100
giant stars in 15 Galactic globular clusters which have good optical
abundance determinations.  The technique is valid for globular cluster giants
with $-1.8<$[Fe/H]$<-0.1$ and $-7<M_{K_{0}}<-4$, and has a typical 
uncertainty of $\pm 0.1$ dex.

Our sample of stars in the different bulge fields has similar colors and
magnitudes to the 
stars analyzed by \citet{ste00}. Figure 3 shows the color-magnitude 
diagram for the globular cluster stars from \citet{ste00} and our sample
of bulge stars with $M_{K_{0}}>-7$. The scatter seen in globular cluster stars
is real and arises from sequences of different metallicities, where
bluer cluster stars are more metal poor. 
The bulge stars appear in a line because of the dereddening technique,
where we force the stars to lie on the BW giant branch.
Figure 4 compares the three spectral indices (EW(Na), EW(Ca), and EW(CO)) 
with dereddened color, $(J-K)_{0}$ for globular cluster and 
bulge stars. 
Note that there is considerable overlap of the two populations although the
globular cluster stars extend to lower values of equivalent widths while the 
bulge stars go to higher values. These differences most likely reflect 
differences in the [Fe/H] distributions of the two populations.

\citet{ste00} calculated two calibrations for globular cluster metallicities.
Solution 1 estimates the metallicity with only the spectral indices EW(Na), 
EW(Ca), and EW(CO).
Solution 2 also incorporates the dereddened $(J-K)$ color and the absolute 
$K-$band magnitude. Figure 5
shows a comparison between results of solution 1 and 2 for the globular cluster
and bulge stars. 
The two solutions yield indistinguishable results for the globular cluster
stars, but for stars in the bulge, solution 2 gives higher 
metallicities for [Fe/H]$>-0.2$. 
Both solutions are extrapolations for [Fe/H] values higher than -0.15 .
Nevertheless, we would like to understand which solution might be better to 
use as an extrapolation to the higher metallicities. 
At higher metallicities the EW(CO) reaches a plateau and
become insensitive to changes in [Fe/H] \citep{ste00}. Since solution 1
has a stronger dependence on the EW(CO) than solution 2, solution 1 is 
expected to be less sensitive to changes in [Fe/H] at higher metallicities. 
The analysis of \citet{ste00} also shows that at higher metallicities 
$M_{K_{0}}$ accounts for more and more of the variation in [Fe/H].
For this reason we feel that solution 2 is a better indicator of metallicity, 
and is the one we applied to our sample of bulge stars.

We applied solution 2 to the individual stars of our sample with $M_{K_{0}} 
\geq -7$, corresponding to the brightest cluster stars. 
If we consider a typical bulge star of $M_{K_{0}} = -6.5$, $(J-K)_{0}$ = 1.1,
EW(Na) = 4.0 \AA, EW(Ca) = 3.0 \AA, EW(CO) = 21.9 \AA, the total 
uncertainties for the equivalent widths (see section 3.1), and the
scatter in the photometric indices due to differential reddening and
dispersion through the line of sight (see section 3.2), we compute a 
typical error in [Fe/H] of $\pm$0.12 dex for individual stars in the g fields
and BW
and $\pm$0.23 dex for individual stars in the c fields. The typical error in 
[Fe/H] is almost doubled for the stars in the c fields, because differential
reddening is higher in these very low latitude fields and the uncertainty in 
the $(J-K)_{0}$ color becomes important. 

We compute a mean value of [Fe/H] for each field by averaging the results of
the individual stars. The mean [Fe/H], the standard deviation and the error in
the mean for each field are listed in Table 6. The error in the mean is the
standard deviation divided by the square root of the number of stars in
each field.

\section{Discussion}

\subsection{Comparison to Slope of Giant Branch method.}

\citet{fro99} used the slope of the giant branch (GB) to estimate the mean 
metallicities of the same c and g fields of our sample. 
For BW we used the slope of the GB result from \citet{tie95}.
In Figure 6 we have plotted slope of the GB results against ours. 
The agreement is very good in all the fields, except for g3-1.3. 
The mean average difference of both techniques is $-0.03 \pm 0.15$ dex, 
entirely consistent with the combined uncertainties of the two techniques.

\subsection{Metallicity Gradients in the Inner Bulge.}

We first explore the possible existence of a metallicity gradient along the 
major-axis
of the inner bulge including all the fields with galactic latitude,
$b=-1.3^{\circ}$.
Figure 7 shows our results for g0-1.3, g1-1.3, g1-1.3, g2-1.3, and g4-1.3
fields plotted against galactic longitude, $l$. The line is
an error weighted least-squares fit to the points.
We find that there might be a small metallicity gradient along the major axis
since the slope of the line is 0.017 $\pm$ 0.011 dex/degree.
In Figure 7, we have also plotted the metallicity gradient along the
major axis obtained by \citet{fro99} (dashed-line).
Note that their result and ours are very close.

Next we explore the existence of a metallicity gradient along the minor-axis
of the inner bulge including all the fields with galactic longitude, $l \sim
0^{\circ}$. 
Figure 8 shows our results for c, g0-1.3, g0-1.8, g0-2.3, g0-2.8, and 
BW fields plotted against galactic latitude, $b$. The line is
an error weighted least-squares fit to the points. 
There is no evidence for a metallicity gradient along the minor axis;
the slope of the fit is -0.012 $\pm$ 0.027 dex/deg. This result seems to be 
in disagreement with earlier results, in which a metallicity gradient along
the minor axis ranges from -0.04 dex/deg \citep{fro90} to -0.09 dex/deg 
\citep{tys91}.
However, if we consider only the metallicities obtained by \citet{fro99} for
the c, g0-1.3, g0-1.8, g0-2.3, g0-2.8, and BW fields we obtain a fit with a 
slope of 0.001 $\pm$ 0.021, in close agreement with our spectroscopic result.
The minor axis metallicity gradient found in earlier studies arises when 
fields at higher galactic latitude are also included. 
In Figure 8, we have plotted as a dashed line the metallicity gradient along the
minor axis obtained by \citet{fro99} including all fields with 
$l \leq 10.5^{\circ}$.

In Figure 9, we have plotted the location of the observed fields with 
respect to a 3.5$\mu$m COBE/DIRBE outline of the Galactic bulge \citep{wei94}.
When only fields located inside the COBE/DIRBE outline are considered,
no metallicity gradient is found. 
The metallicity gradient arises only when fields located
outside the COBE/DIRBE outline ($R > $ 0.6 kpc) are included. 
Metallicity gradients in the galactic bulge have recently been predicted by
the theoretical models of \citet{mol00}.
Moll\'{a} et al. present a multiphase evolution model which assumes a 
dissipative
collapse of the gas from a protogalaxy or halo to form the bulge and the 
disk. They predict a metallicity gradient of -0.4 dex/kpc in the bulge
region 0.5 $ \leq R \leq $ 1.5 kpc, which is in good agreement with the
metallicity gradient found by \citet{fro99}. 
But, Moll\'{a} et al. also predict a steeper gradient, -0.8 dex/kpc, in the 
inner bulge at $ R < $ 0.5 kpc,  which is not observed in our data or in the
\citet{fro99} data. 
We find the metallicity gradient becomes flat at the scale height were the 
infrared light becomes dominant in the Galactic bulge ($R < $ 0.6 kpc). 
\citet{mol00} assume that a core population, which is metal rich and supported
by rotation, dominates the stellar population of the inner bulge. 
The existence of such a metal rich population in the inner bulge is not 
supported
by recent measurements of stellar iron abundances in the Galactic Center
by \citet{ram00}, who found [Fe/H] near solar for a sample of late supergiant
and giant stars. 

\subsection{[Fe/H] metallicity distribution}

In section 4.1 we found that the spread in EW(Na, Ca, CO) at a given magnitude
was consistently higher than expected from measurement uncertainties alone.
A likely explanation for the observed spread in the equivalent width values is
that it arises from an intrinsic spread in [Fe/H] for the stars.

We compute the [Fe/H] metallicity distribution considering all stars in our
sample with $M_{K_{0}} \geq -7$ but excluding the c field stars, because
of their large individual uncertainties in [Fe/H], and including stars
along the major axis, for a total of 72 stars.
The mean [Fe/H] for the inner bulge is -0.21 dex with a full width 
dispersion of 0.30 dex.
Since the average error of our [Fe/H] results is $\pm$0.12 per star (see 
section 4.2), the dispersion observed in the metallicity distribution 
is real.
These values are consistent with theoretical results from \citet{mol00},
who predict a mean [Fe/H] of -0.20 with a dispersion of 0.40 dex for
one bulge population recipe.
If we consider a typical bulge star of $M_{K_{0}} = -6.5$, $(J-K)_{0}$ = 1.1,
EW(Na) = 4.0 \AA, EW(Ca) = 3.0 \AA, EW(CO) = 21.9 \AA, compute [Fe/H]
using solution 2 of \citet{ste00}, and determine the difference in [Fe/H] adding and
subtracting the scatter in the equivalent widths due to metallicity (see Sec.
4.1), we obtain a difference of $\pm$ 0.26. This number is very similar to
the dispersion of the [Fe/H] distribution. We conclude that the scatter seen
in the equivalent widths is real and can be explained by the dispersion
observed in the [Fe/H] distribution.

We now compare the metallicity distribution of our sample of 72 stars in the
inner bulge with the metallicity distribution derived for BW by \cite{sad96} 
in Figure 10.
Their mean [Fe/H] for 262 stars in BW is -0.15 dex with a dispersion of 
0.44 dex. 
This is quite similar to our mean of -0.21 dex with a dispersion of 0.30 dex.

\section{Conclusions.}

We present $K-$band spectra of giant stars in fields interior to $-$4 degrees
and as close as 0.2 degrees of the Galactic Center. We measure equivalent 
widths of the strongest features present in the $K-$band spectra, EW(Na),
EW(Ca), and EW(CO), and also dereddened photometric indices $M_{K_{0}}$ and
$(J-K)_{0}$. We use these indices to compute [Fe/H] for the individual 
stars, using the calibration derived for globular clusters by \citet{ste00}.
The mean [Fe/H] for each field is in good agreement with the results 
obtained with the slope of the giant branch method \citep{fro99}.
We find no evidence for a metallicity gradient along the minor or major
axis of the bulge  for $ R < $ 0.6 kpc. 
We also show that metallicity gradients found in earlier works only
arise when fields located at larger galactic radii are included. Those 
higher galactic radii fields are located outside the infrared bulge defined
by the COBE/DIRBE outline.
We compute the [Fe/H] distribution for the inner bulge, finding a mean value of
-0.21 dex with a full width dispersion of 0.30 dex, which are very similar to 
the mean and width of the BW's [Fe/H] distribution from \citet{sad96} and 
to the theoretical distribution of a bulge formed by dissipative collapse 
\citep{mol00}.

\acknowledgments
S.V.R. gratefully acknowledges support from a Gemini Fellowship 
(grant \# GF-1003-97 from the Association of Universities for Research in 
Astronomy, Inc., under NSF cooperative agreement AST-8947990 and from 
Fundaci\'on Andes under project C-12984), 
and from an Ohio State Presidential Fellowship.
We thank the CTIO staff for helpful support.
J.A.F. thanks the former director of the Carnegie Observatories, Dr. Leonard
Searle, for a Visiting Research Associateship without which this program 
could not have gotten started. 
Finally, J.A.F. notes that NSF declined to provide support for
this research program.


\clearpage

\begin{figure}
\epsfxsize=6.0truein
\epsfbox{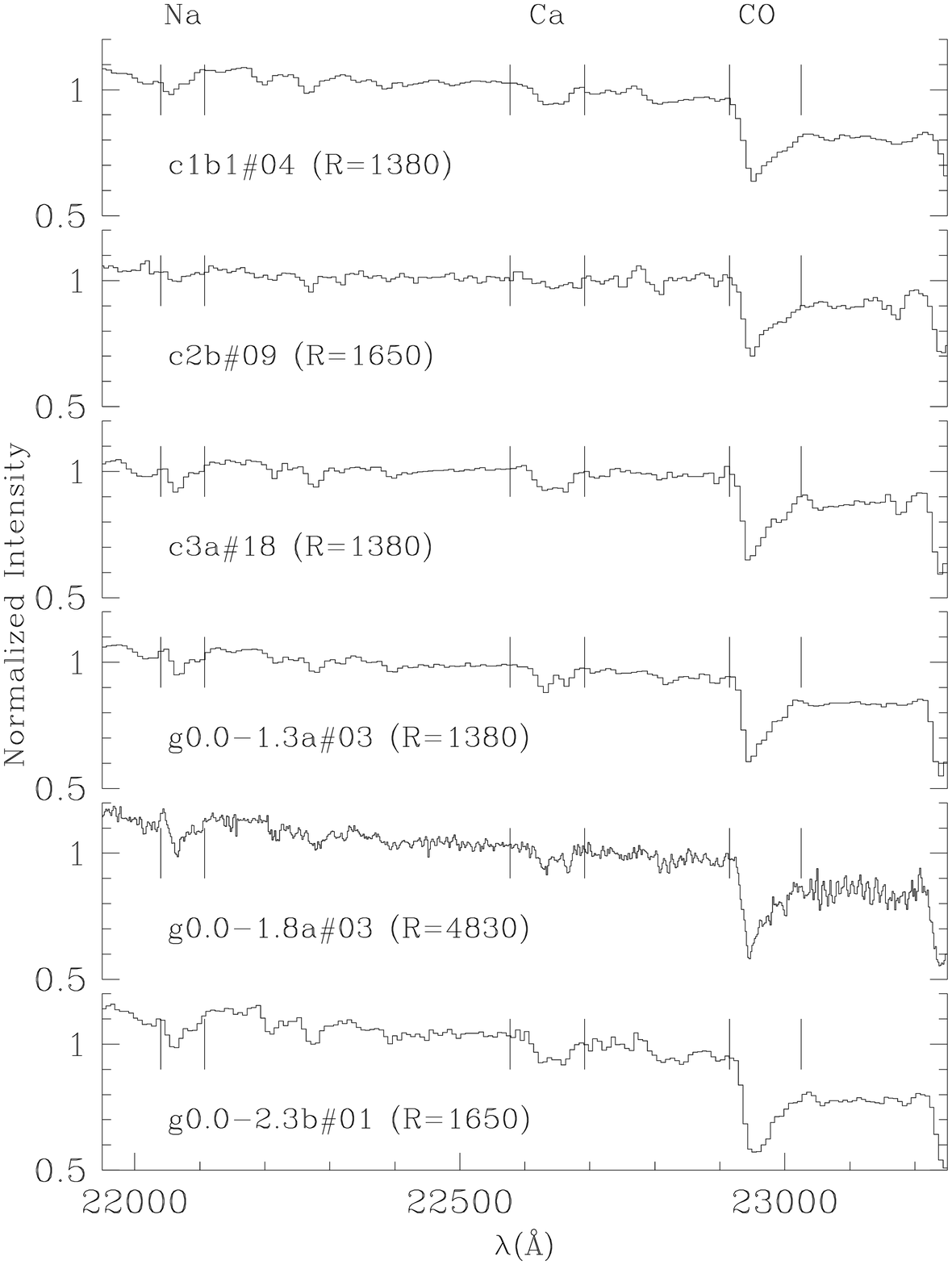}
\caption{Spectra of the brightest stars at each field
presented as normalized flux ($F_{\lambda}$).
Each star is identified below its spectrum.
The vertical bars show the limits of the feature bands.
\label{fig1}}
\end{figure}

\begin{figure}
\epsfxsize=6.0truein 
\epsfbox{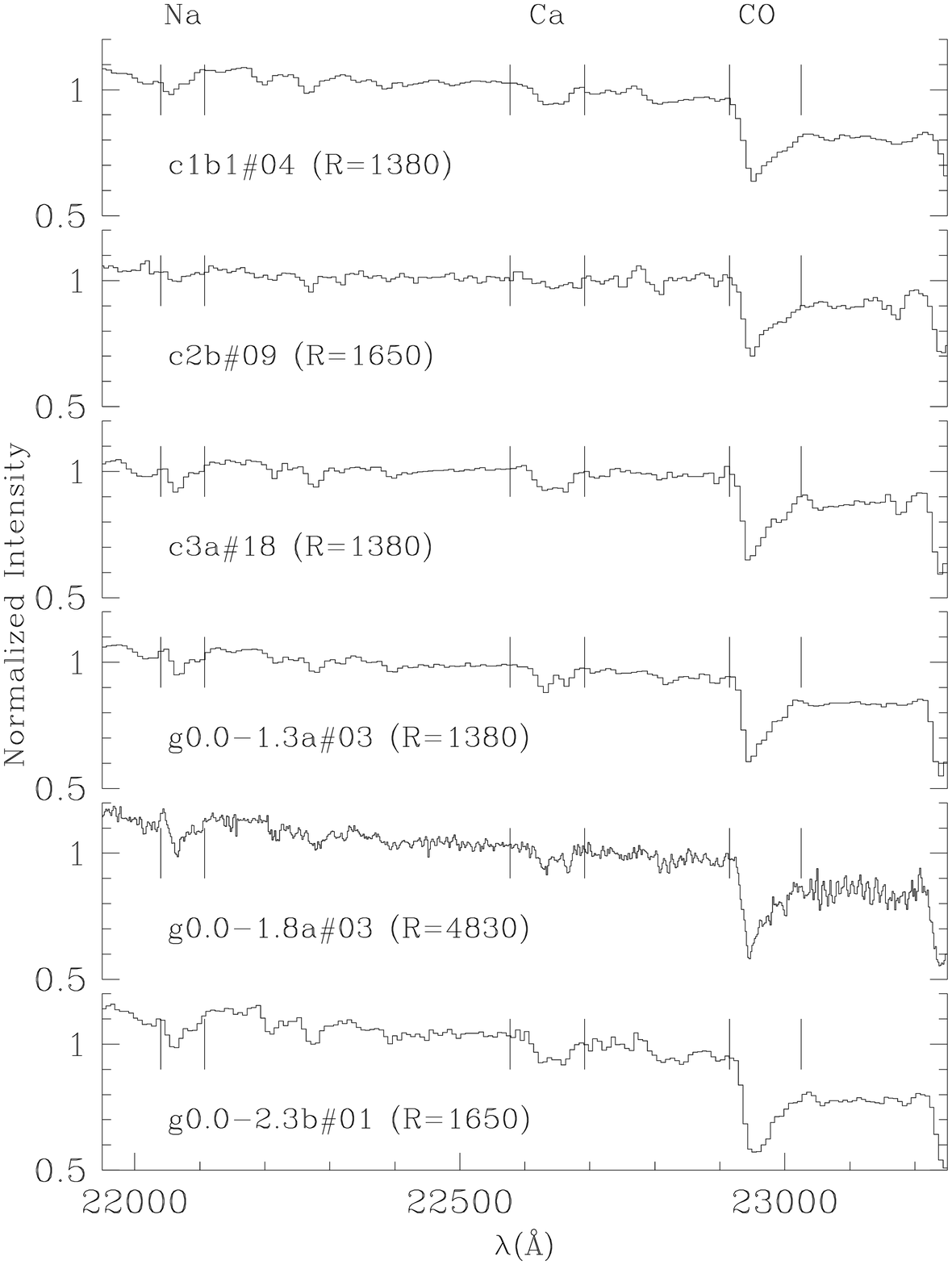}
\end{figure}

\begin{figure}
\epsfxsize=6.0truein 
\epsfbox{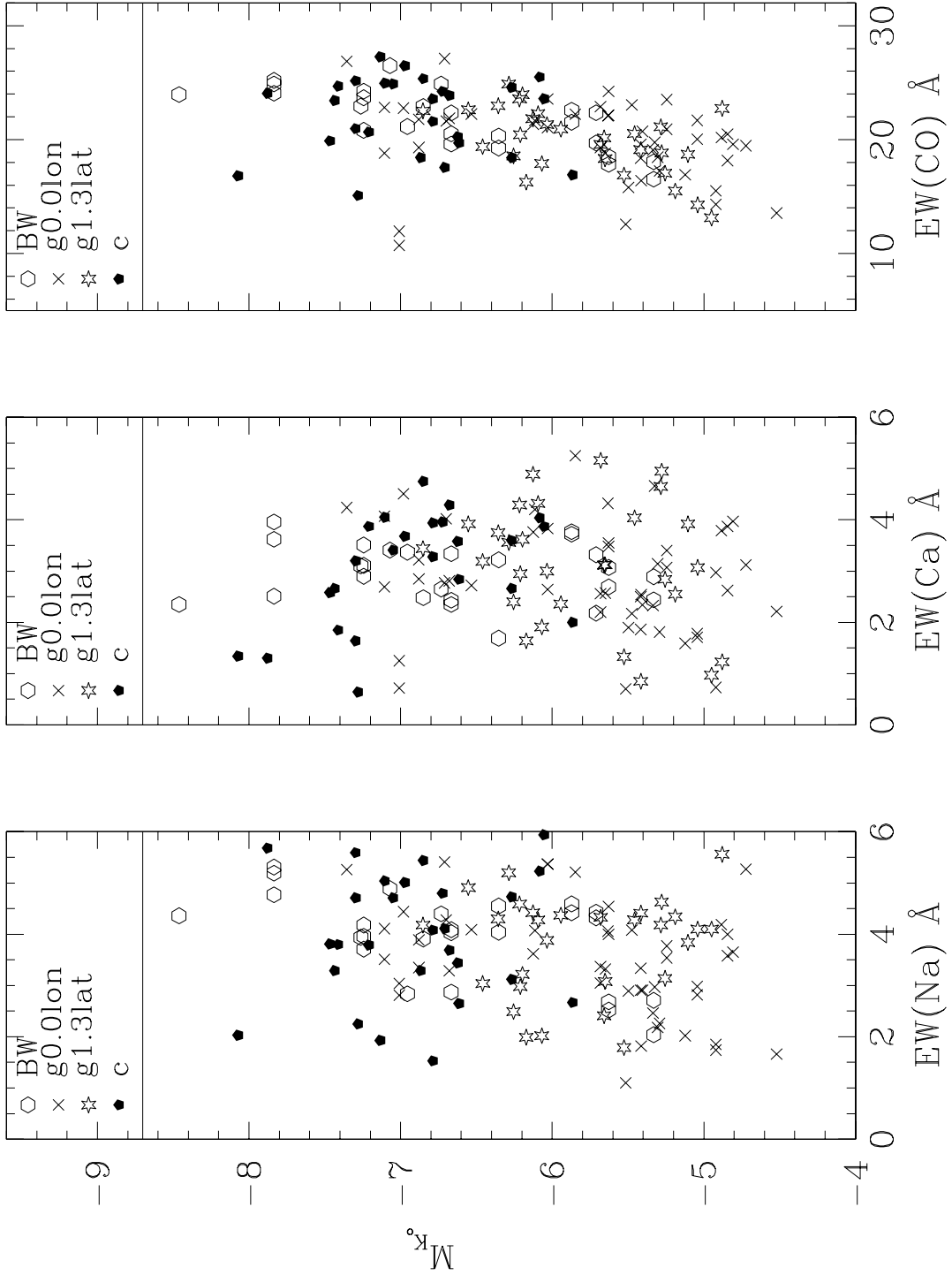}
\caption{Equivalent widths of the spectral features Na, Ca, and CO
plotted against absolute magnitude, $M_{K_{0}}$.
\label{fig2}}
\end{figure}

\begin{figure}
\epsfxsize=6.0truein
\epsfbox{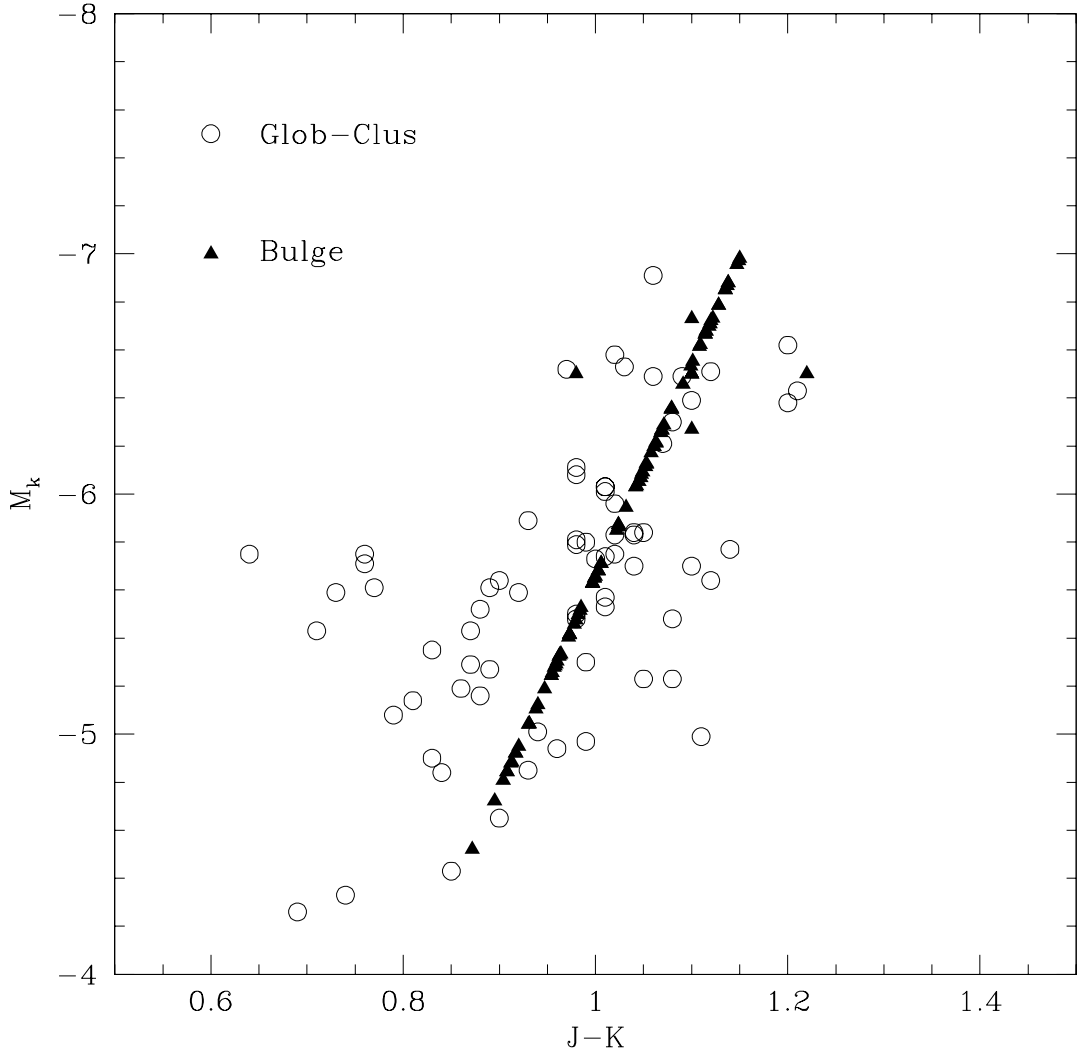}
\caption{Color-Magnitude Diagram of bulge stars from this work 
($filled-triangles$) and globular cluster stars from \citet{ste00}
($open-circles$).
\label{fig3}}
\end{figure}

\begin{figure}
\epsfxsize=6.0truein
\epsfbox{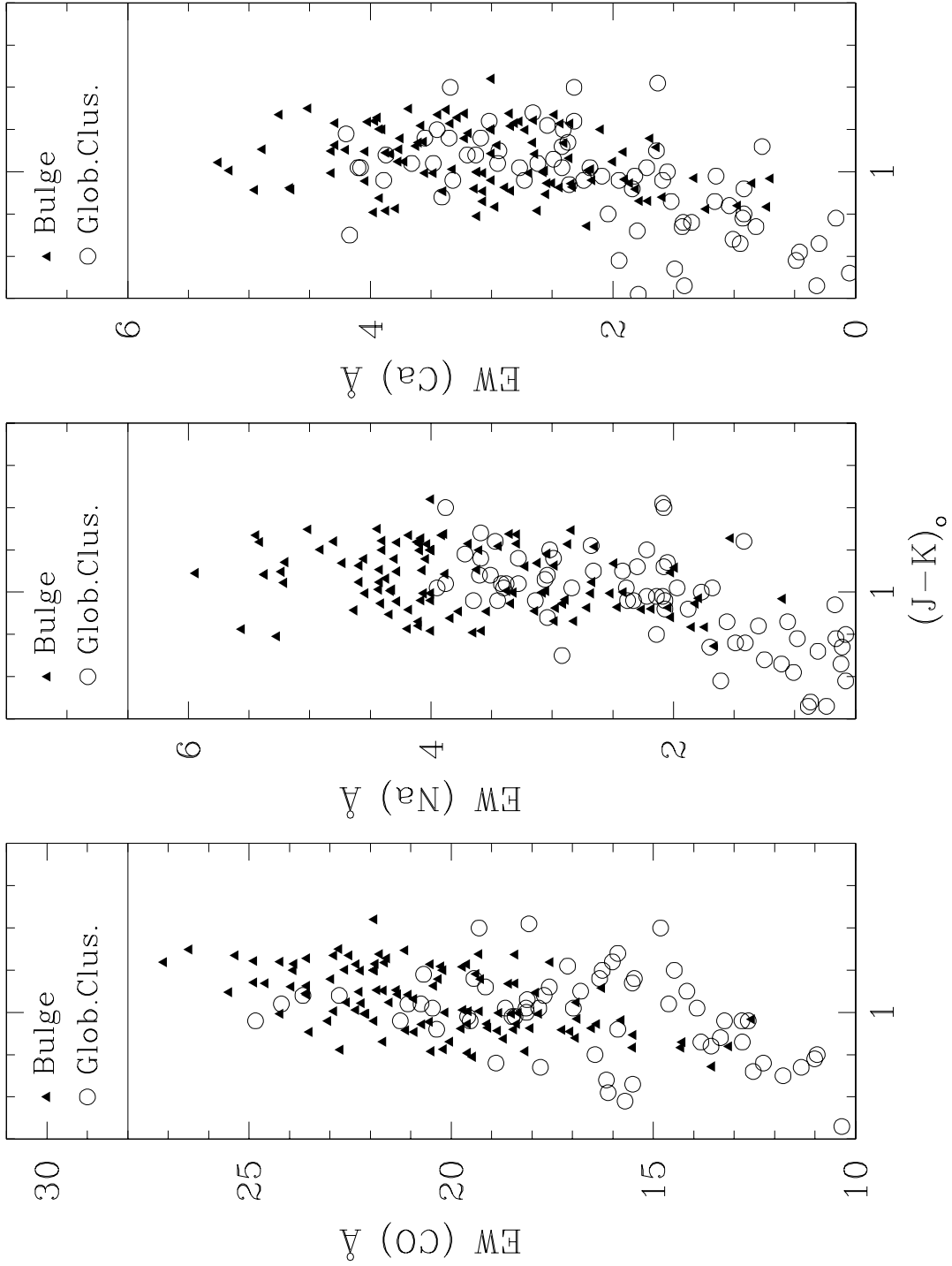}
\caption{Equivalent widths of the spectral features Na, Ca, and CO
plotted against color $(J-K)_{0}$. Bulge stars ($filled-triangles$)
and globular cluster stars ($open-circles$) have similar spectral and
photometric parameters.
\label{fig4}}
\end{figure}

\begin{figure}
\epsfxsize=6.0truein
\epsfbox{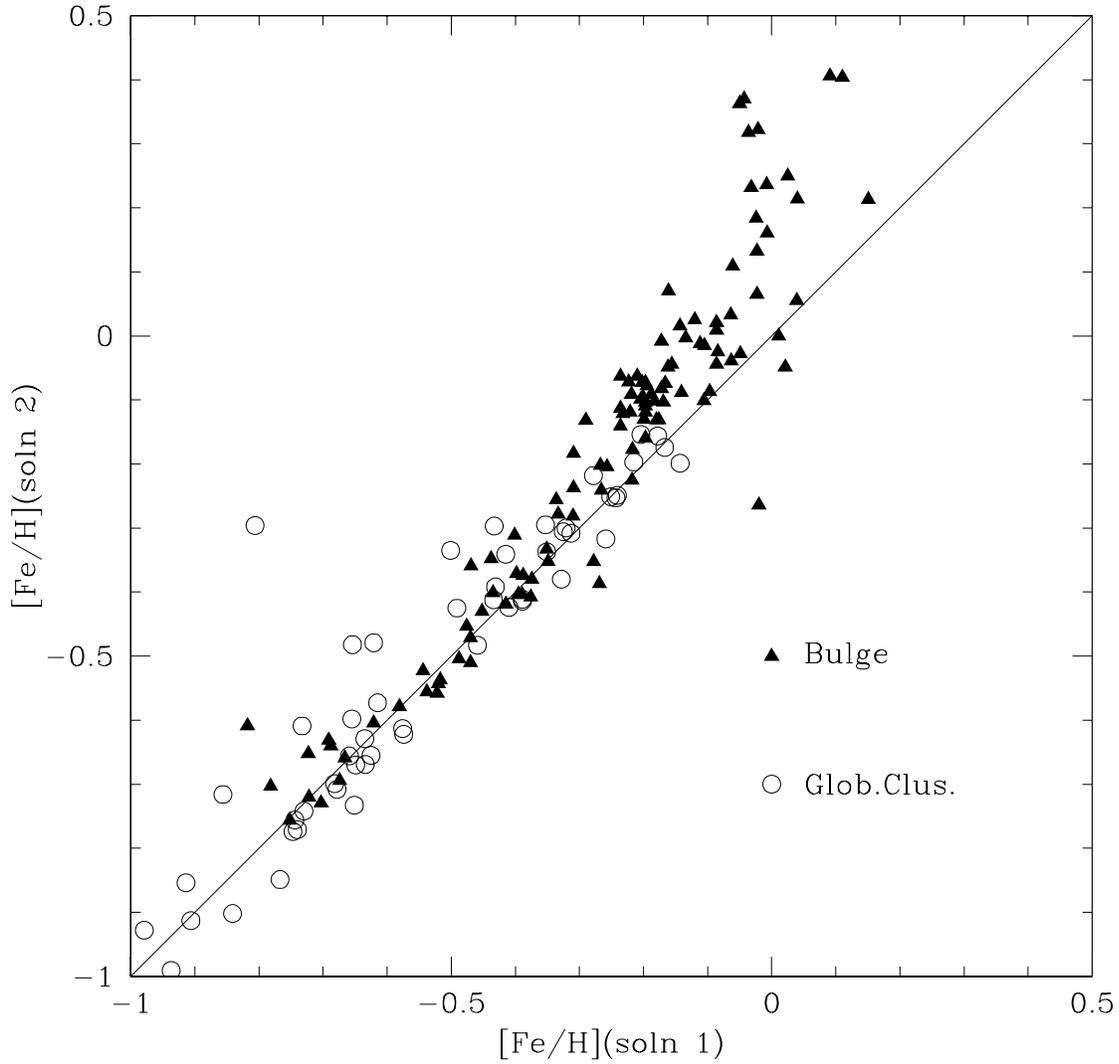}
\caption{Metallicity from solution 1 and from solution 2 applied to
bulge stars ($filled-triangles$) and globular cluster stars ($open-circles$).
\label{fig5}}
\end{figure}

\begin{figure}
\epsfxsize=6.0truein
\epsfbox{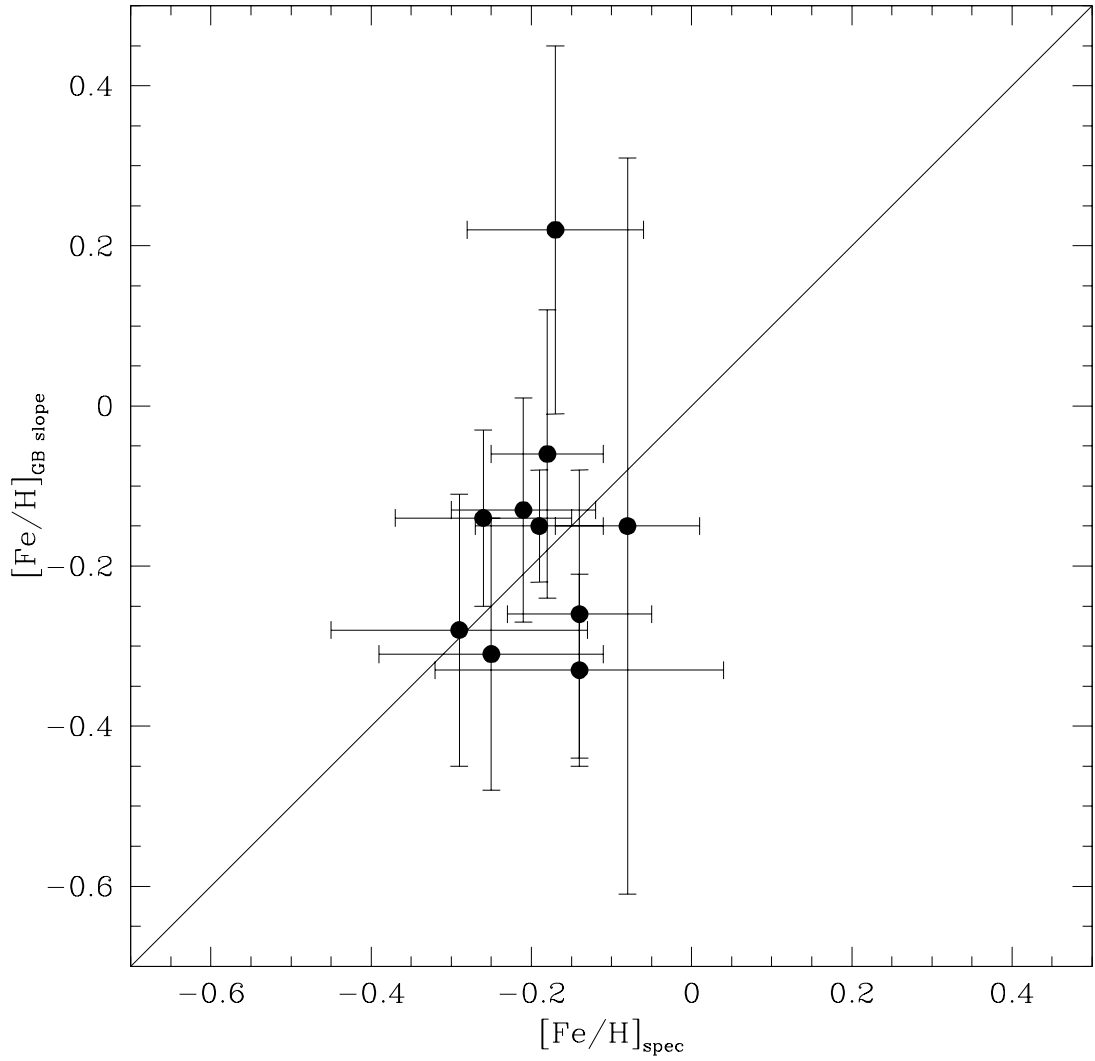}
\caption{[Fe/H] from our spectroscopic calibration plotted against 
[Fe/H] from the slope of the giant branch (GB) from \citet{fro99}.
\label{fig6}}
\end{figure}

\begin{figure}
\epsfxsize=6.0truein
\epsfbox{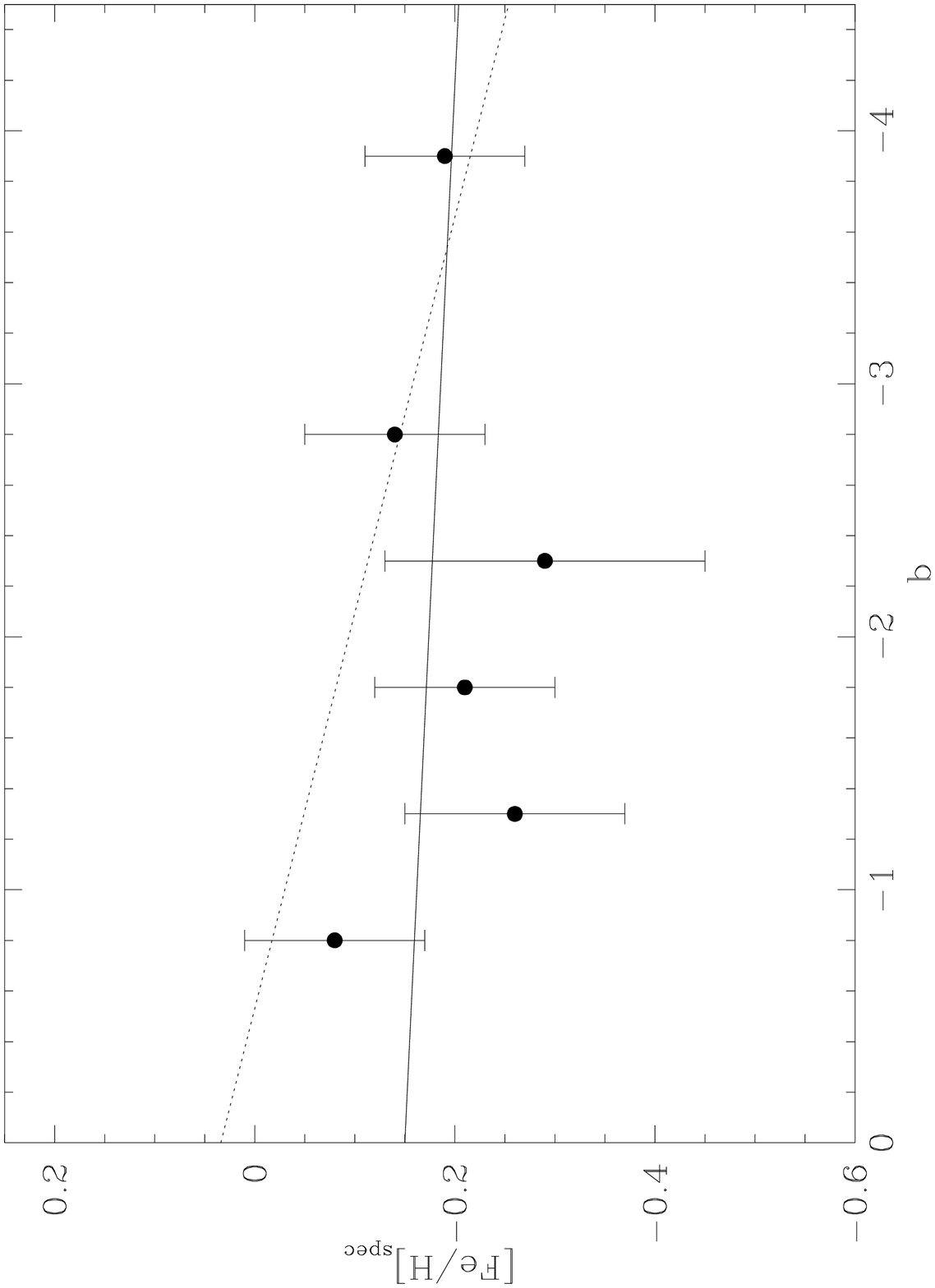}
\caption{[Fe/H] from our spectroscopic calibration plotted against
galactic longitude, $b$ for fields in the inner bulge with galactic latitude
b=$1.3^{\circ}$. The line is an least-squares fit to the points.
The slope of the line is 0.012 $\pm$ 0.018 dex/degree.
\label{fig7}}
\end{figure}

\begin{figure}
\epsfxsize=6.0truein
\epsfbox{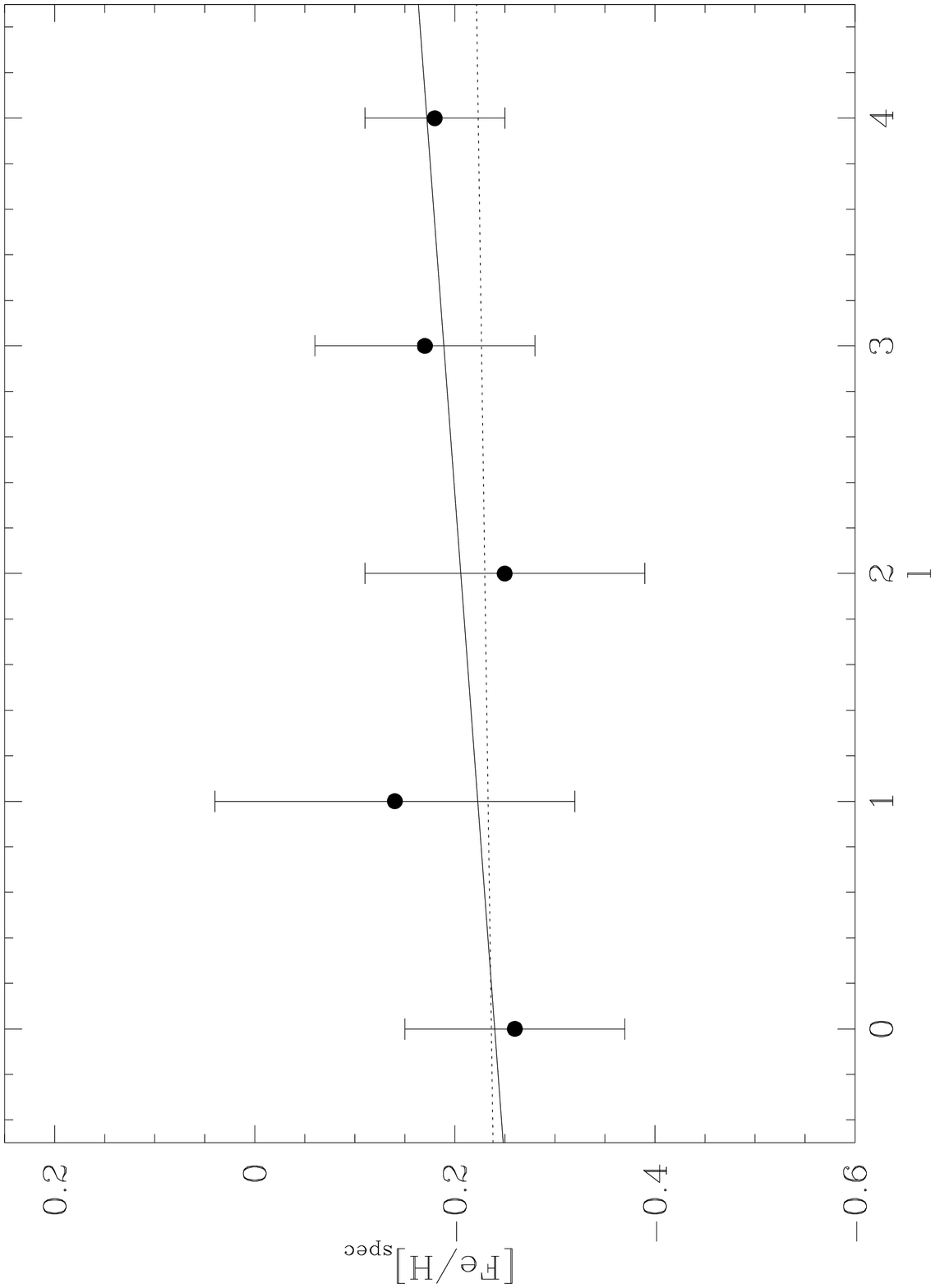}
\caption{[Fe/H] from our spectroscopic calibration plotted against
galactic latitude, $b$ for fields in the inner bulge with galactic longitude,
$l \sim 0^{\circ}$. The line is an least-squares fit to the points. 
The slope of the line is -0.07 $\pm$ 0.05 dex/degree.
\label{fig8}}
\end{figure}

\begin{figure}
\epsfxsize=6.0truein
\epsfbox{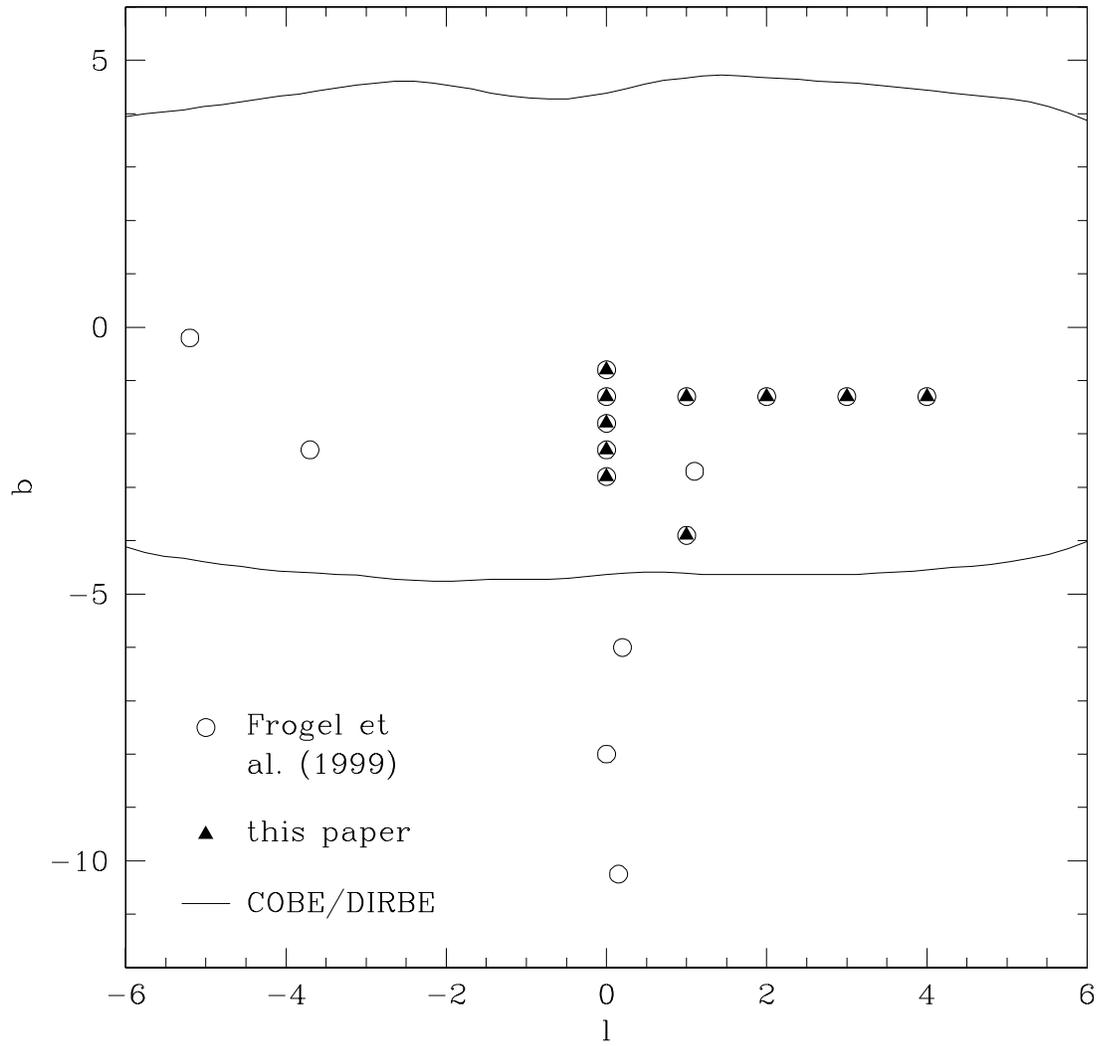}
\caption{Location of observed fields with respect to the COBE/DIRBE
3.5 $\mu$m outline \citep[$solid$ $line$, ][]{wei94} at 5 MJy ${\rm sr^{-1}}$.
The observed fields are from \citet[][$open-circles$]{fro99} and this work
($filled-triangles$).
\label{fig9}}
\end{figure}

\begin{figure}
\epsfxsize=6.0truein
\epsfbox{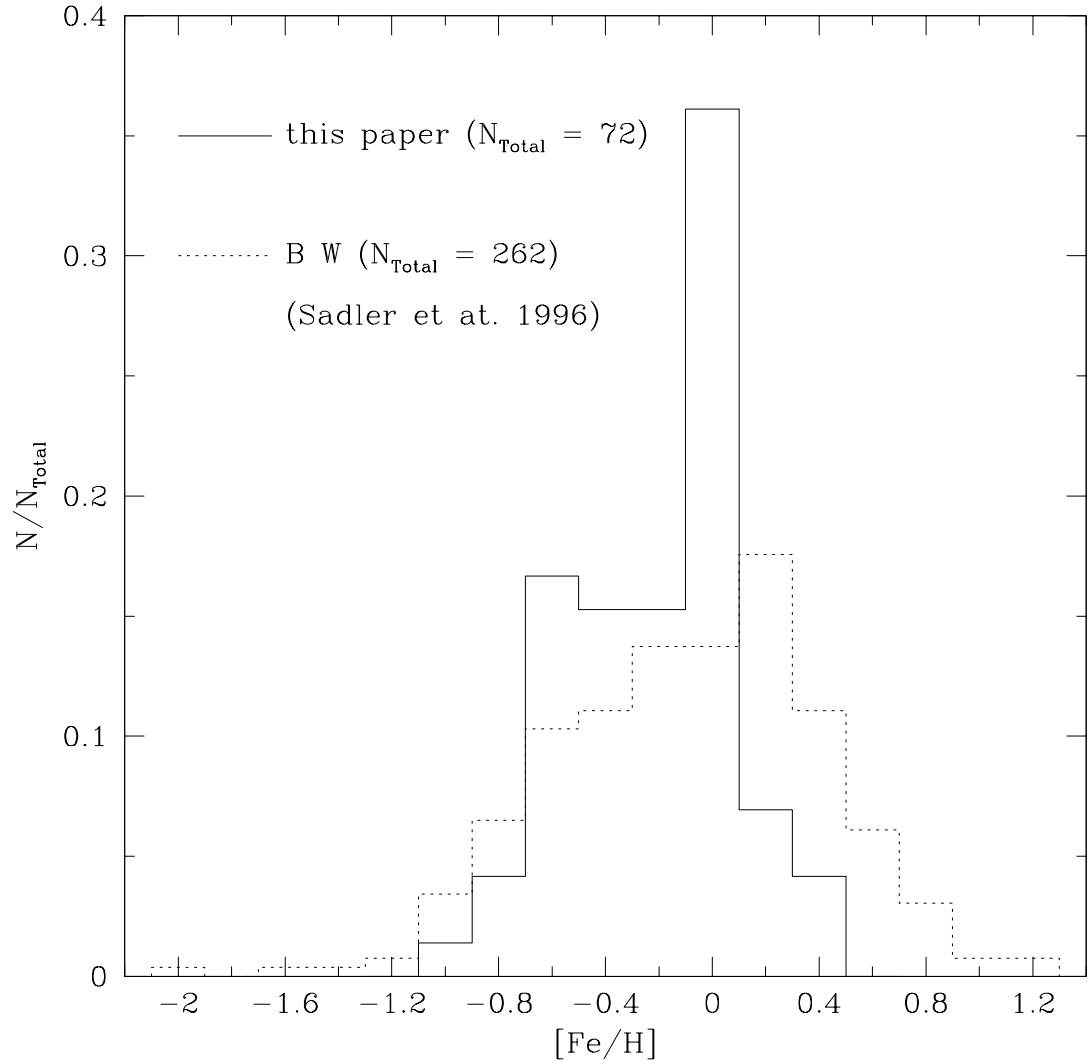}
\caption{Fractional distribution of [Fe/H] for 86 inner bulge stars 
($solid$ $line$) compared to [Fe/H] fractional distribution for 262
Baade's window stars ($dashed$ $line$) from \citet{sad96}.
\label{fig10}}
\end{figure}

\clearpage

\begin{deluxetable}{lcc}
\tablewidth{0pt}
\tablecaption{Observations.}
\tablehead{\colhead{Date} & \colhead{Instrument} & 
\colhead{$\lambda / \Delta \lambda $}} 
\startdata
1993 August 26,27,28     &  OSIRIS  & 1380         \\
1994 July 23,24,25,27,28 &  OSIRIS  & 1380         \\
1995 July 19,20,21,22    &  IRS     & 1650 \& 4830 \\
\enddata
\end{deluxetable}

\clearpage

\begin{table}
\caption{Observed Giant Stars.}
\centerline{
\epsfxsize=7.0truein
\epsfbox[50 150 580 700]{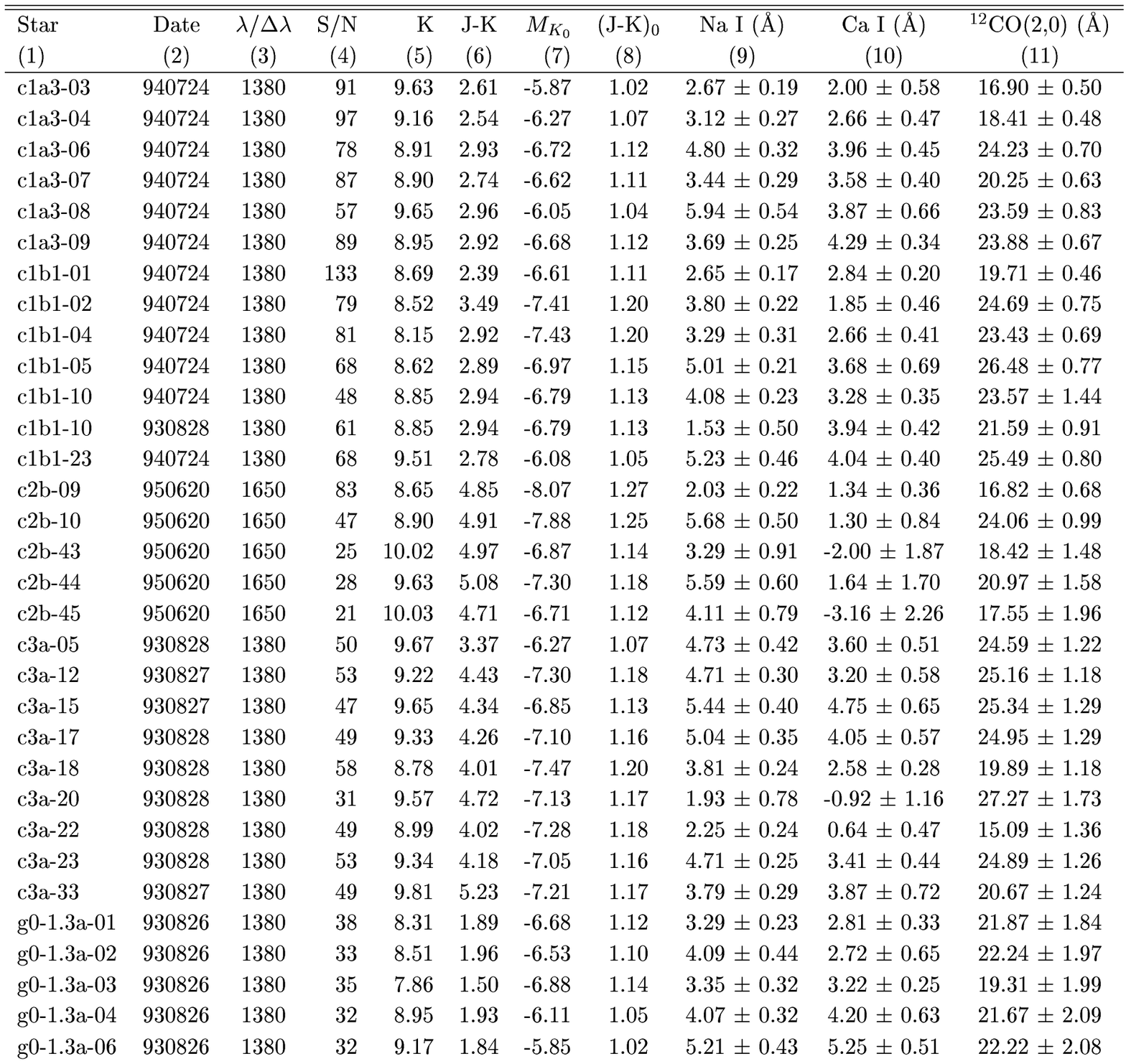}
}
\end{table}

\clearpage

\begin{table}
\centerline{
\epsfxsize=7.0truein
\epsfbox[50 150 580 700]{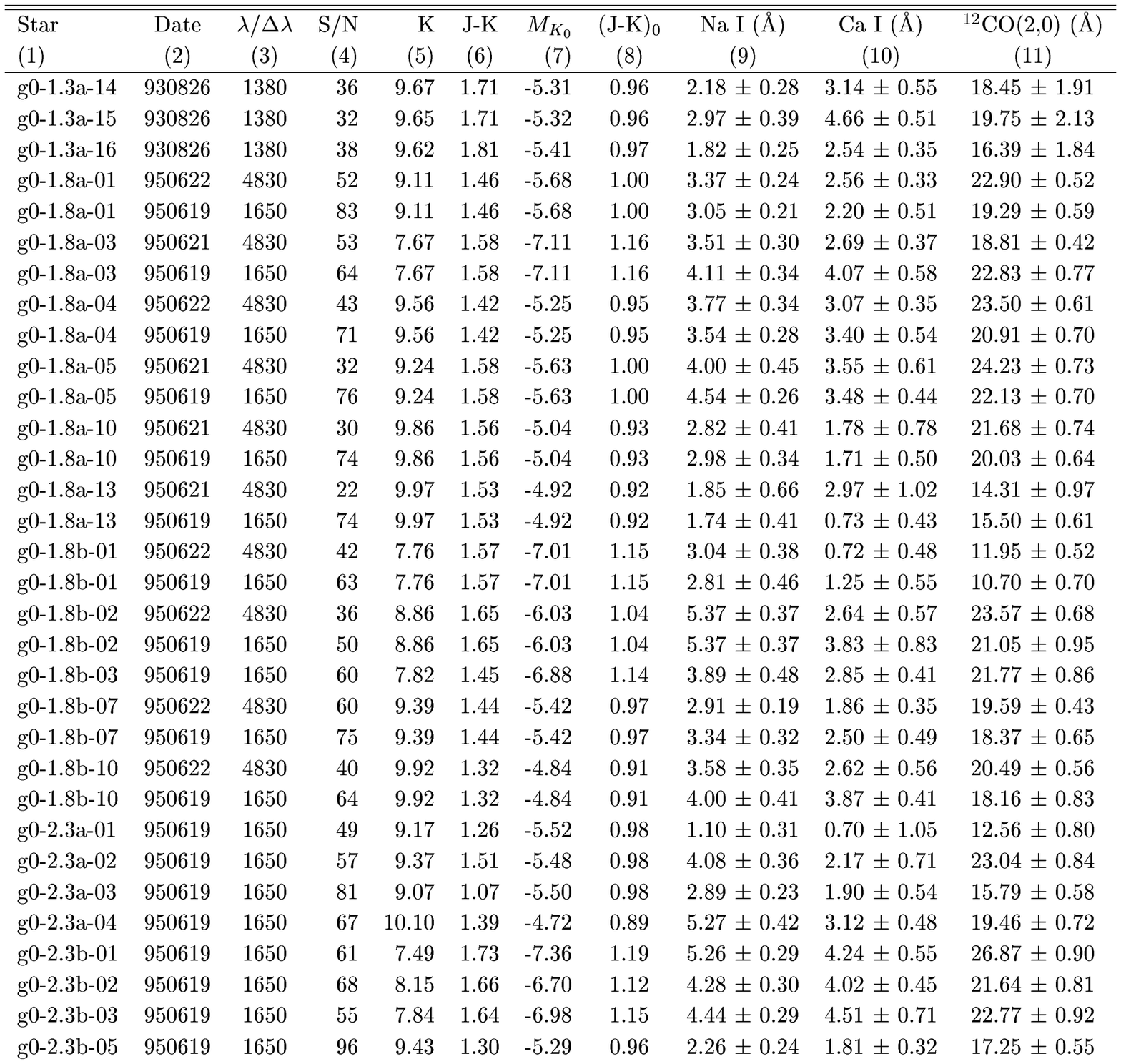}
}
\end{table}

\clearpage

\begin{table}
\centerline{
\epsfxsize=7.0truein
\epsfbox[50 150 580 700]{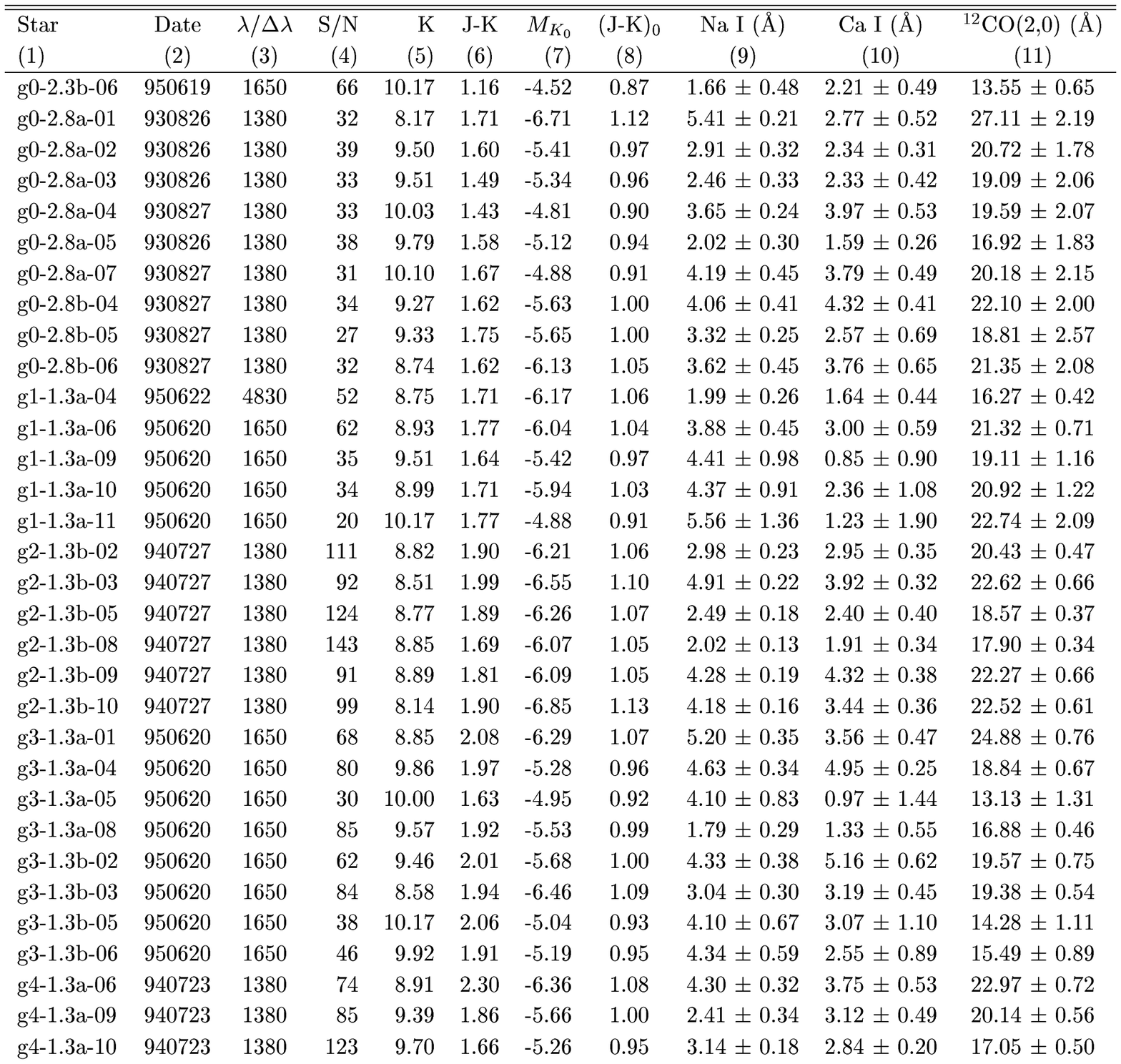}
}
\end{table}

\clearpage

\begin{table}
\centerline{
\epsfxsize=7.0truein
\epsfbox[50 150 580 700]{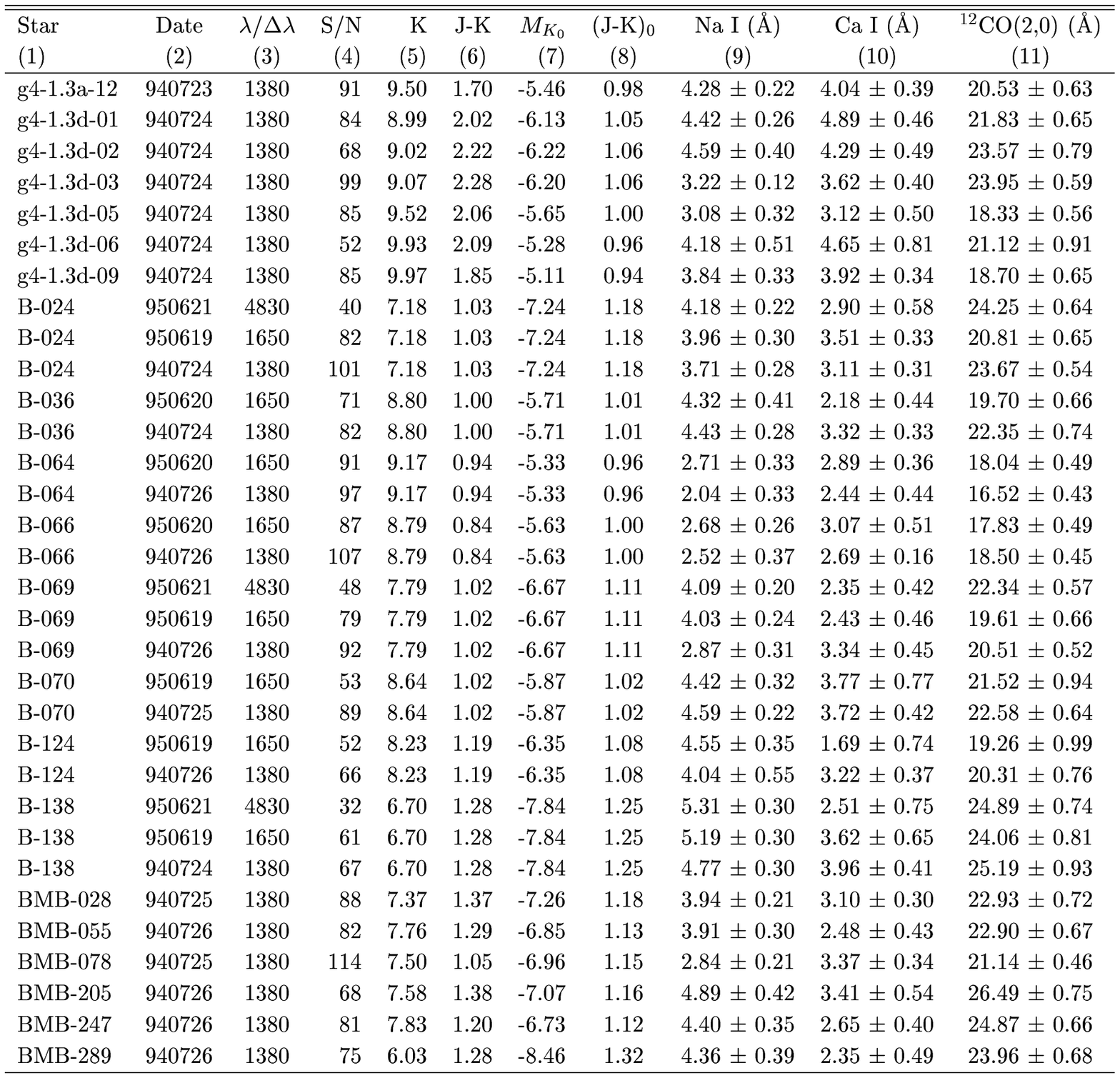}
}
\end{table}

\clearpage

\begin{deluxetable}{lc}
\tablewidth{0pt}
\tablecaption{Definitions of band edges for continuum and features.}
\tablehead{\colhead{Feature Name} & \colhead{band edges ($\mu$m)}}
\startdata
Na continuum \# 1             & 2.191 $-$ 2.197  \\
Na continuum \# 2             & 2.213 $-$ 2.217  \\
Na I feature                  & 2.204 $-$ 2.211  \\
                              &                  \\
Ca continuum \# 1             & 2.245 $-$ 2.256  \\
Ca continuum \# 2             & 2.270 $-$ 2.272  \\
Ca I feature                  & 2.258 $-$ 2.269  \\ 
                              &                  \\
$ ^{12}$CO(2,0) continuum \#1 & 2.190 $-$ 2.201  \\
$ ^{12}$CO(2,0) continuum \#2 & 2.211 $-$ 2.222  \\
$ ^{12}$CO(2,0) continuum \#3 & 2.233 $-$ 2.260  \\    
$ ^{12}$CO(2,0) continuum \#4 & 2.268 $-$ 2.280  \\
$ ^{12}$CO(2,0) continuum \#5 & 2.286 $-$ 2.291  \\
$ ^{12}$CO(2,0) bandhead      & 2.292 $-$ 2.303  \\
\enddata
\end{deluxetable}

\clearpage

\begin{deluxetable}{lccccc}
\tablewidth{0pt}
\tablecaption{Average differences in EW.}
\tablehead{ \colhead{} & 
\multicolumn{2}{c}{$\rm \Delta (IRS_{4830} - IRS_{1650})$} &
\multicolumn{2}{c}{$\rm \Delta (IRS_{1650} - OSIRIS_{1380})$} & \colhead{} \\
\colhead{} & \multicolumn{2}{c}{N=13} & \multicolumn{2}{c}{N=8}
& \colhead{N=150} \\
\colhead{} & \colhead{Mean}$\Delta$  & \colhead{Standard Deviation}  
           & \colhead{Mean}$\Delta$  & \colhead{Standard Deviation} 
& \colhead{$<$Formal error$>$} }
\startdata
EW(Na) & -0.06 & 0.32 & -0.36 & 0.43 & 0.34 \\
EW(Ca) & -0.33 & 0.96 &  0.33 & 0.78 & 0.54 \\
EW(CO) &  1.50 & 2.10 &  1.10 & 1.30 & 1.04 \\
\enddata
\end{deluxetable}

\clearpage

\begin{deluxetable}{lccc}
\tablewidth{0pt}
\tablecaption{Standard Deviations of Equivalent Widths in two ranges of
$M_{K_{0}}$}
\tablehead{ \colhead{} &
\colhead{$-6 \leq M_{K_{0}} \leq -7 $} &
\colhead{$-5 \leq M_{K_{0}} \leq -6 $} &
\colhead{Total Uncertainties} \\
\colhead{EW} & \colhead{$\sigma$} & \colhead{$\sigma$} & \colhead{$\sigma$}}
\startdata
Na & 0.98 & 0.95 & 0.38 \\
Ca & 0.98 & 0.95 & 0.87 \\
CO & 2.40 & 2.52 & 1.70 \\
\enddata
\end{deluxetable}

\clearpage

\begin{deluxetable}{lrccc}
\tablewidth{0pt}
\tablecaption{Metallicity.}
\tablehead{\colhead{Field} & \colhead{N} & \colhead{$<$[Fe/H]$>$} 
& \colhead{$\sigma$} & \colhead{Error in Mean}}
\startdata
c      & 14 & --0.08 & 0.33 & 0.09 \\
g0-1.3 &  8 & --0.26 & 0.28 & 0.11 \\
g0-1.8 &  9 & --0.21 & 0.25 & 0.09 \\
g0-2.3 &  8 & --0.29 & 0.42 & 0.16 \\
g0-2.8 &  9 & --0.14 & 0.25 & 0.09 \\
g1-1.3 &  5 & --0.14 & 0.35 & 0.18 \\
g2-1.3 &  6 & --0.25 & 0.32 & 0.14 \\
g3-1.3 &  8 & --0.17 & 0.28 & 0.11 \\
g4-1.3 & 10 & --0.18 & 0.20 & 0.07 \\
BW     &  9 & --0.19 & 0.21 & 0.07 \\
\enddata
\end{deluxetable}

\end{document}